\documentclass[pra,twocolumn,showpacs,superscriptaddress]{revtex4}
\usepackage{epsfig,graphicx,amssymb,color,bm}
\usepackage{amsmath}
\usepackage[colorlinks, linkcolor=blue, anchorcolor=blue, citecolor=blue, citecolor=blue,unicode=true]{hyperref}
\usepackage{setspace}
\usepackage{url}
\usepackage{hyperref}

\newcommand{\eref}[1]{Eq.~(\ref{#1})}
\newcommand{\fref}[1]{Fig.~\ref{#1}}
\begin{document}
\title{Multi-path photon-phonon converter in optomechanical system at single-quantum level}
\author{Tian-Yi Chen}
\author{Wen-Zhao Zhang}
\author{Ren-Zhou Fang}
\author{Cheng-Zhou Hang}
\author{Ling Zhou}\thanks{zhlhxn@dlut.edu.cn}
\address{School of Physics and Optoelectronic Technology, Dalian University of Technology, Dalian, 116024, People's Republic of China}
\begin{abstract}
Based on photon-phonon nonlinear interaction, a scheme is proposed to realize a controllable multi-path photon-phonon converter at single-quantum level in a composed quadratically coupled optomechanical system.
Considering the realization of the scheme, an associated mechanical oscillator is introduced to enhance the effective nonlinear effect.
Thus, the single-photon state can be converted to the phonon state with high fidelity even under the current experimental condition that the single-photon coupling rate is much smaller than mechanical frequency ($g\ll\omega_m$).
The state transfer protocols and their transfer fidelity are discussed both analytically and numerically.
A multi-path photon-phonon converter is designed, by combining the optomechanical system with low frequency resonators, which can be controlled by experimentally adjustable parameters.
This work provides us a potential platform for quantum state transfer and quantum information processing.
\end{abstract}
\maketitle
\section{Introduction}
The radiation pressure in optomechanical system provides an excellent interaction between optical cavity mode and microcosmic or macroscopic mechanical mode \cite{RevModPhys.86.1391,PhysRevLett.116.163602}.
In addition to some promising applications in fundamental physics research \cite{srep.6.23678,PhysRevA.94.012334}, macroscopic mechanical oscillators cooling \cite{PhysRevA.93.063853,PhysRevA.91.033818,PhysRevA.91.063815}, weak force sensing \cite{PhysRevLett.114.113601,PhysRevLett.113.151102,PhysRevLett.97.133601,NJP.18.103047} and quantum information processing \cite{JPB.48.015502,PhysRevLett.109.013603,PhysRevE.93.062221}, optomechanical devices provide an outstanding characteristic to transduce an input states with a given frequency into an output states with another frequency while preserving quantum properties at the same time.
The quantum-state and entanglement can be converted from light to macroscopic oscillators \cite{PhysRevA.68.013808,PhysRevA.86.021801} via optomechanical systems in optical regime.
In electro-opto-mechanical system, electrical and optical quantum states can be stored and transferred into mechanical resonators \cite{PhysRevA.87.053818,ANDP:ANDP201400116} thus the system is able to serve as a microwave quantum-illumination device \cite{PhysRevLett.114.080503}.
Since the interaction of beam splitter in optomechanical system, which constructs the converted effect, relies on linearization condition which requires large existence of photons in optomechanical cavity, it is impossible to achieve the single photon-phonon conversion in optomechanical system under this condition.
On the other hand, many proposal of convertors, transferring few photon state to different frequency electromagnetic wave state based on quantum nonlinearity \cite{PhysRevA.89.053813,srep.3.3555,OL.30.2375}, have been proposed, including four-wave mixing media converter \cite{OL.30.2375}, single-photon frequency conversion in a Sagnac interferometer \cite{srep.3.3555}.
There is a great challenge in converter when huge frequency difference exists between input state and output state, such as optical mode and mechanical mode, due to the requirement of strong nonlinear interaction in this kind of scheme.
Under current experimental parameters region, how to enhance the effective nonlinearity and how to employ their nonlinearity to perform quantum information processing deserve our investigation.

In this paper, we put forward a scheme to enhance the cross-Kerr nonlinearity in a quadratically coupled optomechanical system.
Considering the realization, we use an auxiliary mechanical oscillators to enhance the quantum nonlinear effects in the system, and thus achieve an ultra-strong cross-Kerr nonlinearity ($g_{eff}/\omega_{eff} \gg 1$).
By combining single bit operations in optical mode and mechanical mode, we can implement a photon-phonon converter at single-quantum level.
Then we construct a multi-path photon-phonon converter by extending the dimension of system, which can be controlled by experimentally adjustable parameters.

\section{model}
\begin{figure}[b]
  \centering
  \includegraphics[width=8cm]{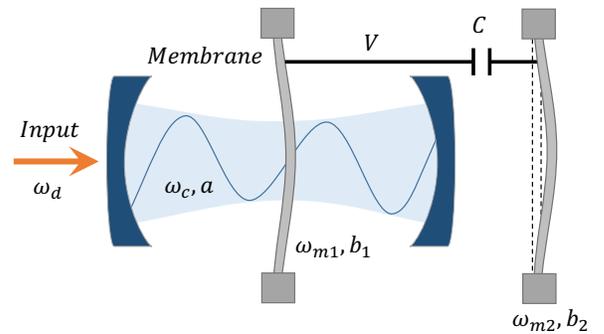}\\
  \caption{The quadratically coupled optomechanical system consists of a membrane in the middle of the cavity. The membrane is interacted to a low frequency resonator with the coupling strength $V$ via a capacitance $C$.}\label{fig1}
\end{figure}
We consider a quadratically coupled optomechanical system with a membrane in the middle of a Fabry-P\'{e}rot cavity, which is coupled with a low frequency mechanical oscillator, as the figure shown in \fref{fig1}.
The mechanical displacement of the membrane quadratically couples to the cavity photon number.
The interaction between two oscillators can be realized by a resonator interacting with a transmission line resonator through the medium of capacitance \cite{NJP.10.115001}, or by using the geometrically interconnecting \cite{nphys.9.480}.
The Hamiltonian, in which $\hbar=1$, of the system is $H=H_{sys}+H_{d}$,
\begin{eqnarray}\label{hami}
  H_{sys} &=& \omega_c a^{\dag}a-ga^{\dag}a(b_1+b_1^{\dag})^2\nonumber\\
  &&+\sum_{j=1,2}(\omega_{mj}b_j^{\dag}b_j+Vb_{j}^{\dag}b_{3-j}),\\
  H_{d}&=&\varepsilon (a^{\dag}e^{-i\omega_dt}+ae^{i\omega_dt}),
\end{eqnarray}
where $a^{\dag}(a), b_{1}^{\dag}(b_{1})$ and $b_2^{\dag}(b_2)$ are the creation (annihilation) operators of the F-P cavity, the mechanical membrane and the auxiliary mechanical oscillator, respectively. $\omega_c$, $\omega_{m1}$ and $\omega_{m2}$ are the resonant frequency of them. The second term in $H_{sys}$ describes the quadratic optomechanical coupling between the original cavity and the mechanical membrane with strength $g$.
The last term in $H_{sys}$ describes the free energy and the phonon tunneling coupling between two oscillators with strength $V$ \cite{PhysRevLett.111.103605,PhysRevLett.111.073603}. The input driving of the cavity can be described as $H_{d}=\varepsilon (a^{\dag}e^{-i\omega_dt}+ae^{i\omega_dt})$.
By eliminating the rapid evolution mode $b_1$ due to large frequency $\omega_{m1}$, we obtain the effective interaction between the optomechanical cavity and the auxiliary oscillator mode $b_2$, under the condition $\omega_{m1}\gg\{ \omega_{m2},V,g\}$ (Details are in APPENDIX). The effective Hamiltonian is
\begin{equation}\label{heff}
H_{eff}=\Delta' a^{\dag}a+\omega_{eff}b_2^{\dag}b_2+g_{eff}a^{\dag}ab_2^{\dag}b_2,
\end{equation}
where $\Delta'_{1}=\omega_c-\omega_{d}-g$ denotes the mechanically modulating detuning of the cavity with a driving frequency $\omega_d$. The effective frequency of the mechanical oscillator, coupling strength, and dumpling rates are described by
$\omega_{eff} = \omega_{m2}-\frac{V^2}{\omega_{m1}}$, $g_{eff}=\frac{V^2}{\omega_{m1}^2}2g$, $\gamma_{eff}=\gamma_2+\frac{V^2}{\omega_{m1}^2}\gamma_1$.
We can clearly see the cross-Kerr nonlinear term between cavity mode and mechanical mode $g_{eff}a^{\dag}ab_2^{\dag}b_2$. It is a pivotal effect that can provide a way to preform manipulation between photons and phonons.
Similar with the quantum control schemes based on cross-Kerr nonlinearity \cite{JPB.48.015502,PhysRevA.85.052326}, the key factor of the controlling realization is the weight of nonlinear coupling rate $g_{eff}$ in the Hamiltonian, i.e. $g_{eff}\sim{\Delta',\omega_{eff}},\gamma_{eff}$.
In our system $\Delta'$ is an adjustable parameter which can be controlled by input driving frequency and can be easily reduced.
\begin{figure}
  \centering
  \includegraphics[width=8.5cm]{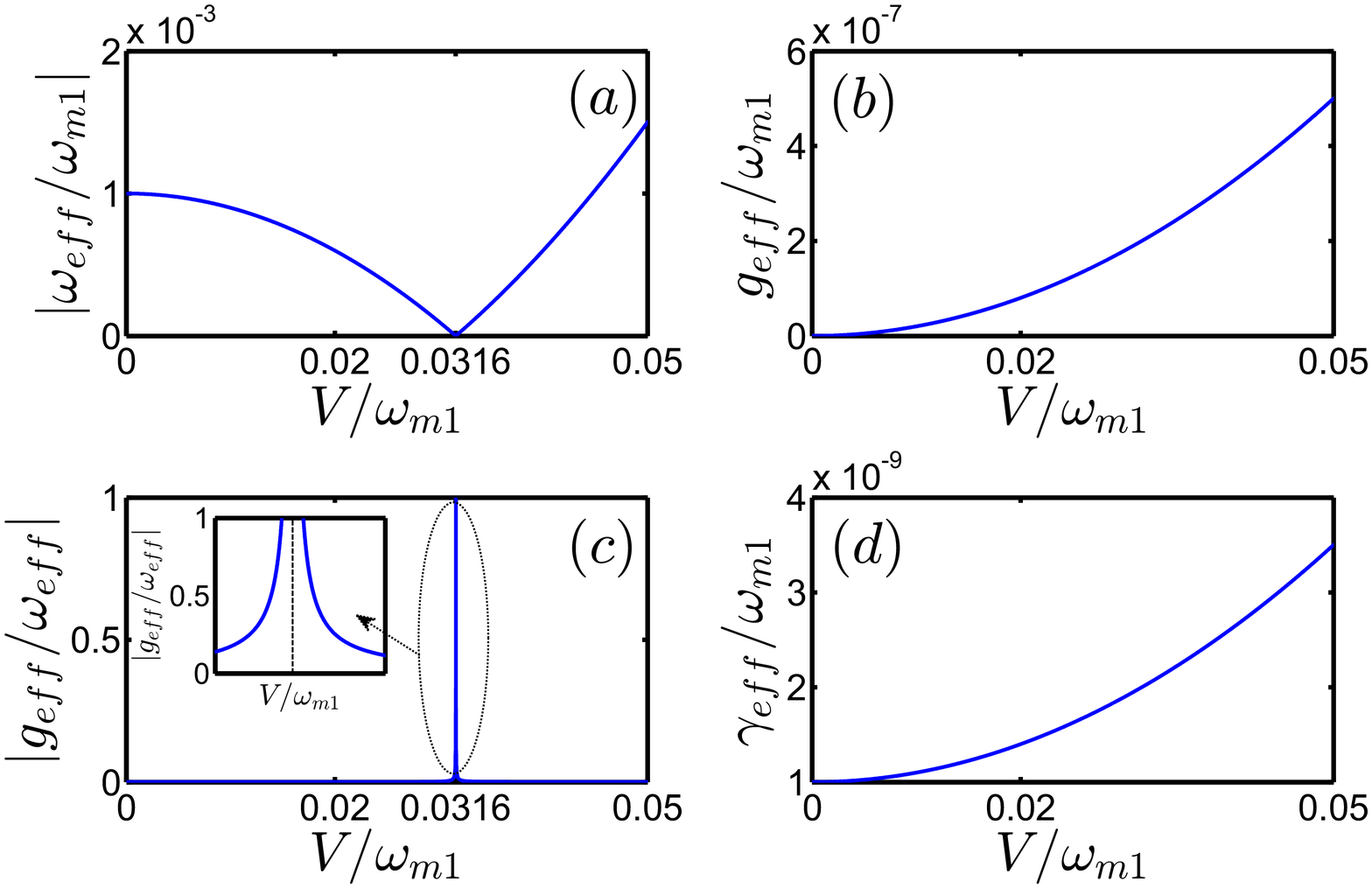}\\
  \caption{The effective frequency of the mechanical oscillator $\omega_{eff}$, effective coupling strength $g_{eff}$, the ratio of $|\frac{g_{eff}}{\omega_{eff}}|$ and effective damping rate $\gamma_{eff}$ is a function of mechanical coupling strength $V$.
  Other parameters are $g/\omega_{m1}=10^{-4}$, $\omega_{m2}/\omega_{m1}=10^{-3}$, $\gamma_{m1}/\omega_{m1}=10^{-6}$, $\gamma_{m2}/\omega_{m2}=10^{-6}$.}\label{fig2}
\end{figure}

Under the condition of $\omega_{m1}\gg V$, we have the effective dispassion rate $\gamma_{eff}\approx \gamma_2$. Here we set $\gamma_2<g_{eff}$.
Now we focus on the effective parameters $\omega_{eff}$, $g_{eff}$ and $\gamma_{eff}$.
As is shown in \fref{fig2}, the effective parameters $g_{eff},\omega_{eff},\gamma_{eff}$ reduce compared with the original ones.
As is shown in \fref{fig2}a, there is a minimal value of the effective mechanical frequency $|\omega_{eff}|=0$ at the specific value of $V=\sqrt{\omega_{m1}\omega_{m2}}$.
Meanwhile, the effective coupling rate $g_{eff}$ and damping rate $\gamma_{eff}$ increase with the mechanical coupling rate $V$ rising, which is shown in \fref{fig2}b and \fref{fig2}d.
As is shown in \fref{fig2}c, we plot the ratio between the effective coupling rate and the effective mechanical frequency $|g_{eff}/\omega_{eff}|$. There is a discontinuity point tending to infinity at a specific value of mechanical coupling rate $V=\sqrt{\omega_{m1}\omega_{m2}}$.
Thus we can let $|g_{eff}/\omega_{eff}|\gg1$ by adjusting the coupling rate under the condition $V\ll\omega_{m1}$.
So it is possible for us to achieve an ultra-strong cross-Kerr nonlinearity in the system.

\section{Photon-phonon control phase-flip gate}
\begin{figure}
  \centering
  \includegraphics[width=8cm]{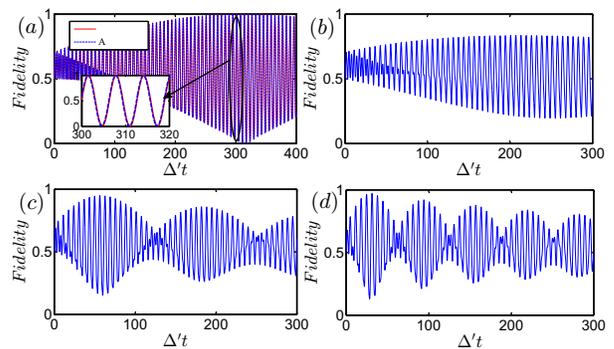}\\
  \caption{(a) The comparison of analytical and numerical solution. The mechanical coupling rate $V/\omega_{m1}=3.131\times 10^{-2}$.
  (b), (c) and (d) denote the fidelity as a function of time $t$ with different mechanical coupling strength $V$.
  (b) The mechanical coupling rate $V/\omega_{m1}=3.131\times10^{-2}$. The corresponding ratio $|\frac{g_{eff}}{\omega_{eff}}|=0.01$, and the maximal fidelity $F_{max}\approx 0.83$.
  (c) The mechanical coupling rate $V/\omega_{m1}=3.156\times10^{-2}$. The corresponding ratio $|\frac{g_{eff}}{\omega_{eff}}|=0.05$, and the maximal fidelity $F_{max}\approx 0.94$.
  (d) The mechanical coupling rate $V/\omega_{m1}=3.159\times10^{-3}$. The corresponding ratio  $|\frac{g_{eff}}{\omega_{eff}}|=0.1$, and the maximal fidelity $F_{max}\approx 0.97$.
  The optical dissipation rate $\kappa/g=0.2$, the thermal excitation number $n_{th}=1$.
  Other parameters are the same with \fref{fig2}.}\label{fig3}
\end{figure}

Now we show that the composed optomechanical system can work as a photon-phonon control phase-flip gate based on the cross-Kerr nonlinearity shown in the third term of the effective Hamiltonian in \eref{heff}.
We use the ground- (excited-) state of photon and phonon to denote the logical states $|0\rangle(|1\rangle)$ of signal mode and control mode, respectively.
The unknown signal qubit inputs via the optical cavity. The arbitrary initial state of the system can be describes as
\begin{eqnarray}
  |\psi\rangle &=& \alpha|0{\rangle}_{c}|0{\rangle}_{m}+\beta|0{\rangle}_{c}|1{\rangle}_{m}
  +\gamma|1{\rangle}_{c}|0{\rangle}_{m}+\delta|1{\rangle}_{c}|1{\rangle}_{m},\nonumber\\
\end{eqnarray}
where $|\alpha|^2+|\beta|^2+|\gamma|^2+|\delta|^2=1$. $|0{\rangle}_{c}(|1{\rangle}_{c})$ means that no (one) photon is in the cavity while $|0{\rangle}_{m}(|1{\rangle}_{m})$ means the ground state (the first excited state) of the oscillator.
After we accomplish the optomechanical control phase-flip gate (CPFG), the target state should be
\begin{eqnarray}
  |\Phi\rangle &=& \alpha|0{\rangle}_{c}|0{\rangle}_{m}+\beta|0{\rangle}_{c}|1{\rangle}_{m}
  +\gamma|1{\rangle}_{c}|0{\rangle}_{m}-\delta|1{\rangle}_{c}|1{\rangle}_{m}.\nonumber\\
\end{eqnarray}
If we use the effective Hamiltonian and ignore the cavity's decay and the mechanical damping, the finial state is $|\psi_{f}\rangle=e^{-iH_{eff}t}|\psi\rangle$, which can be described as
\begin{eqnarray}
  |\psi_{f}\rangle &=&
  {\alpha}e^{-i\theta_{00}}|0{\rangle}_{c}|0{\rangle}_{m}+{\beta}e^{-i\theta_{01}}|0{\rangle}_{c}|1{\rangle}_{m} \nonumber \\
  &&+{\gamma}e^{-i\theta_{10}}|1{\rangle}_{c}|0{\rangle}_{m}+{\delta}e^{-i\theta_{11}}|1{\rangle}_{c}|1{\rangle}_{m},
\end{eqnarray}
where $\theta_{00}=0$, $\theta_{01}=\omega_{eff}t$, $\theta_{10}=\Delta't$, $\theta_{11}=(\omega_{eff}+\Delta'-g_{eff})t$.
We define the fidelity $F_{c-p}=|{\langle}\psi_{f}|\Phi\rangle|$ between the final state and the target state is
\begin{eqnarray}
  F_{c-p} &=& |\alpha^{2}e^{-i\theta_{00}}+\beta^{2}e^{-i\theta_{01}}
  +\gamma^{2}e^{-i\theta_{10}}-\delta^{2}e^{-i\theta_{11}}|,\nonumber\\
\end{eqnarray}
If the conditions $\theta_{00}=0$, $\theta_{01}=2n_{1}\pi$, $\theta_{10}=2n_{2}\pi$, $\theta_{11}=(2n_{3}+1)$, $n_i(i=1,2,3)$ can be any real number, are satisfied, thus $F_{c-p}=1$ which means the CPFG realized. Then we have the equation sets
\begin{eqnarray}
  \theta_{00} &=& 0 ,\nonumber\\
  \theta_{01} &=&\omega_{eff}t= 2n_{1}\pi,\nonumber\\
  \theta_{10} &=& \Delta't = 2n_{2}\pi,\nonumber\\
  \theta_{11} &=& (\omega_{eff}+\Delta'-g_{eff})t = (2n_{3}+1)\pi,
\end{eqnarray}
and the ratio of the parameters $g_{eff}:\omega_{eff}:\Delta'=n_1:n_2:(n_{1}+n_{2}-n_{3}-1/2)$.
In addition to the analytical solution, we can directly employ the master equation of the system to reconsider CPFG. Including the dissipation of the system, we write the master equation as
\begin{eqnarray}
  \dot{\rho} &=& \frac{i}{\hbar}[\rho,H]+\kappa\mathcal{D}
[a]\rho\\\nonumber &&+\sum_{j=1,2}\gamma_j(n_{thj}+1)\mathcal{D}[b_j]\rho+{\gamma_j}n_{thj}\mathcal{D}[b_j^{\dagger }]\rho,
\end{eqnarray}
where $H$ is the original Hamiltonian, $\kappa$, $\gamma_j$ and $n_{thj}$ are the decay rates of the cavity, mechanical resonator and the thermal occupancy of the mechanical bath respectively.
$\mathcal{D}[o]\rho=o\rho o^{\dagger}-o^{\dagger }o\rho/2 -\rho o^{\dagger }o/2$ is the Lindblad dissipation superoperator.
On this condition, $F_{c-p}=\sqrt{\langle\Phi|\rho|\Phi\rangle}$.
Then we plot the fidelity in both analytical and numerical method in \fref{fig3}a.
As shown in \fref{fig3}a, the CPF gate can be realized in a specific time when the fidelity $F=1$ and the additional phase equals to $(2n+1)\pi$, $n\in \Re$, which is caused by the accumulation of the effect of cross-Kerr nonlinear term $g_{eff}a^{\dag}a b_2^{\dag}b_2$.
Comparing the analytical and numerical solution, we find that the two lines almost coincide. Thus we can safely conclude that the effective Hamiltonian and the analytical solution are correct, except some point with low fidelity due to the approximation in analytical solution.

We notice that, the key point to realize the CPF gate is the term $g_{eff}a^{\dag}a b_2^{\dag}b_2$ in \eref{heff}.
When considering the dissipation, the maximal fidelity decreases due to the accumulation of the dissipation.
Thus, we should reduce this effect to maintain the high fidelity. Here, we minimize the evolution time of the system to achieve this purpose by enlarge the ratio of $|\frac{g_{eff}}{\omega_{eff}}|$.
As shown in \ref{fig3}b, c and d, a larger value of the ratio $|g_{eff}/\omega_{eff}|$ will cause a faster dynamic speed of the CPF gate and also minimize the realization time of the CPF gate, which will reduce the accumulation of the dissipation and obtain a higher fidelity of the CPG gate.
The maximal fidelity with ratio $|\frac{g_{eff}}{\omega_{eff}}|=\{0.01,0.05,0.1\}$ are $F_{max}\approx \{0.83,0.94,0.97\}$, respectively.
If $|g_{eff}/\omega_{eff}|=0.1$, the fidelity will reach 0.97 which means that our model can be used as an optics-mechanical controlled gate.

In this part, based on cross-Kerr nonlinearity effect between cavity and mechanical oscillator, a scheme is proposed to realize a high fidelity controlled-phase gate between photons and phonons under weak coupling regime. This kind of controlled-phase gate is an important quantum device to precess quantum information\cite{JPB.48.015502}.

\section{Single-quantum photon-phonon convertor}
\begin{figure}[h]
  \centering
  \includegraphics[width=8.5cm]{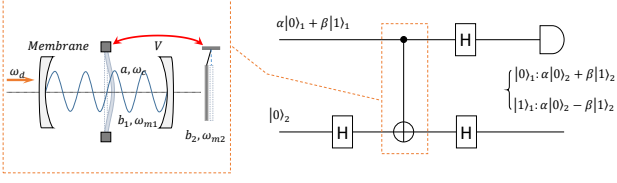}\\
  \caption{Quantum circuit of photon-phonon convertor. After performing single-qubit operation according to the measure result, we can get the state we want in mechanical mode.}\label{fig5}
\end{figure}
By using the character of cavity-oscillator interaction in optomechanical system, we build a link between photons and phonons.
Different from the common method in which the photon state is transferred to phonon with large number of photons in the cavity or strong driving strength under linearized approximation,
we propose a scheme using the cross-Kerr nonlinearity effect to realize a photon-phonon convertor at the single-photon level.
As shown in \fref{fig5}, the quantum circuit denotes a basic process to realize photon-phonon convertor. The photon state can be input through the cavity. After local operators and photon detection we get the required phonon state. The single-qubit code in photons can be easily operated by linear optical device.
The ground- and single- phonon state can be manipulated by film bulk acoustic resonator\cite{PhysRevA.82.032101,nphys.9.480,Nature.464.697}. In \fref{fig5}, the c-phase gate is realized by the optomechanical system with fidelity $F_{max}=0.97$ which we have mentioned in section III.
To transfer an arbitrary optical state $\alpha|0\rangle_1+\beta|0\rangle_1$ to the mechanical oscillator through the convertor, we input the coded state into the cavity while the mechanical oscillator should be cooled into its ground state $|0\rangle_2$.
After performing the gate operator shows in \fref{fig5}, we get the system state,
\begin{eqnarray}
  |0\rangle_1(\alpha|0\rangle_2+\beta|1\rangle_2)+|1\rangle_1(\alpha|0\rangle_2-\beta|1\rangle_2),
\end{eqnarray}
here we should output the optical signal by using the Q-switching when $t=\pi/g_{eff}$ to ensure the high fidelity. Then we detect the output photon from the the cavity. If the photon counting is zero, mechanical oscillator will collapse to the state $\alpha|0\rangle_2+\beta|1\rangle_2$, thus we get the state we want. If the photon counting is one, the mechanical oscillator will collapse to the state $\alpha|0\rangle_2-\beta|1\rangle_2$. Then we just need to performing a $\sigma_z$ operator to get the state we want.
\section{Controllable Multi-path photon-phonon converter}
\begin{figure}
  \vspace{0.1cm}
  \centering
  \includegraphics[width=8cm]{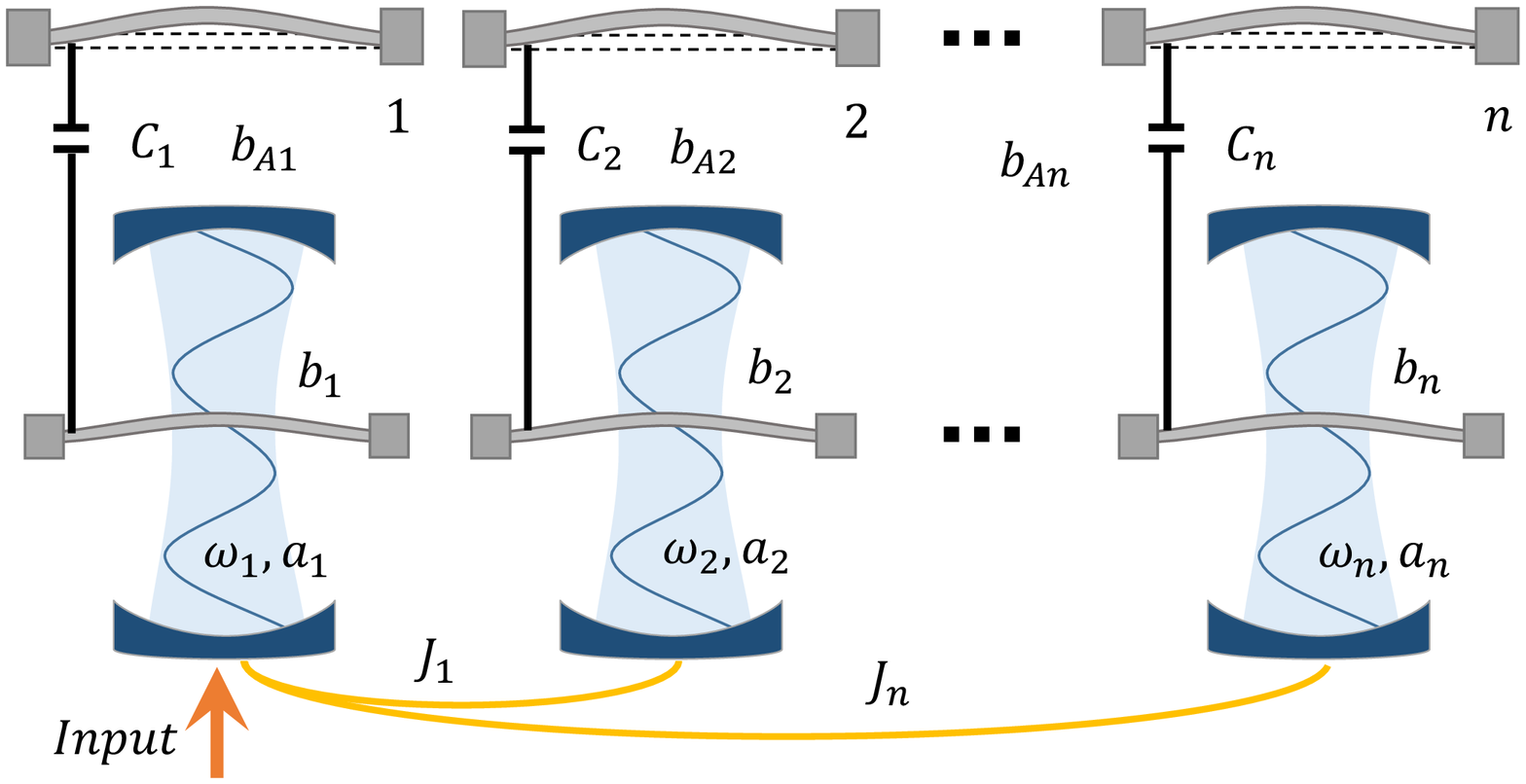}\\
  \includegraphics[width=7cm]{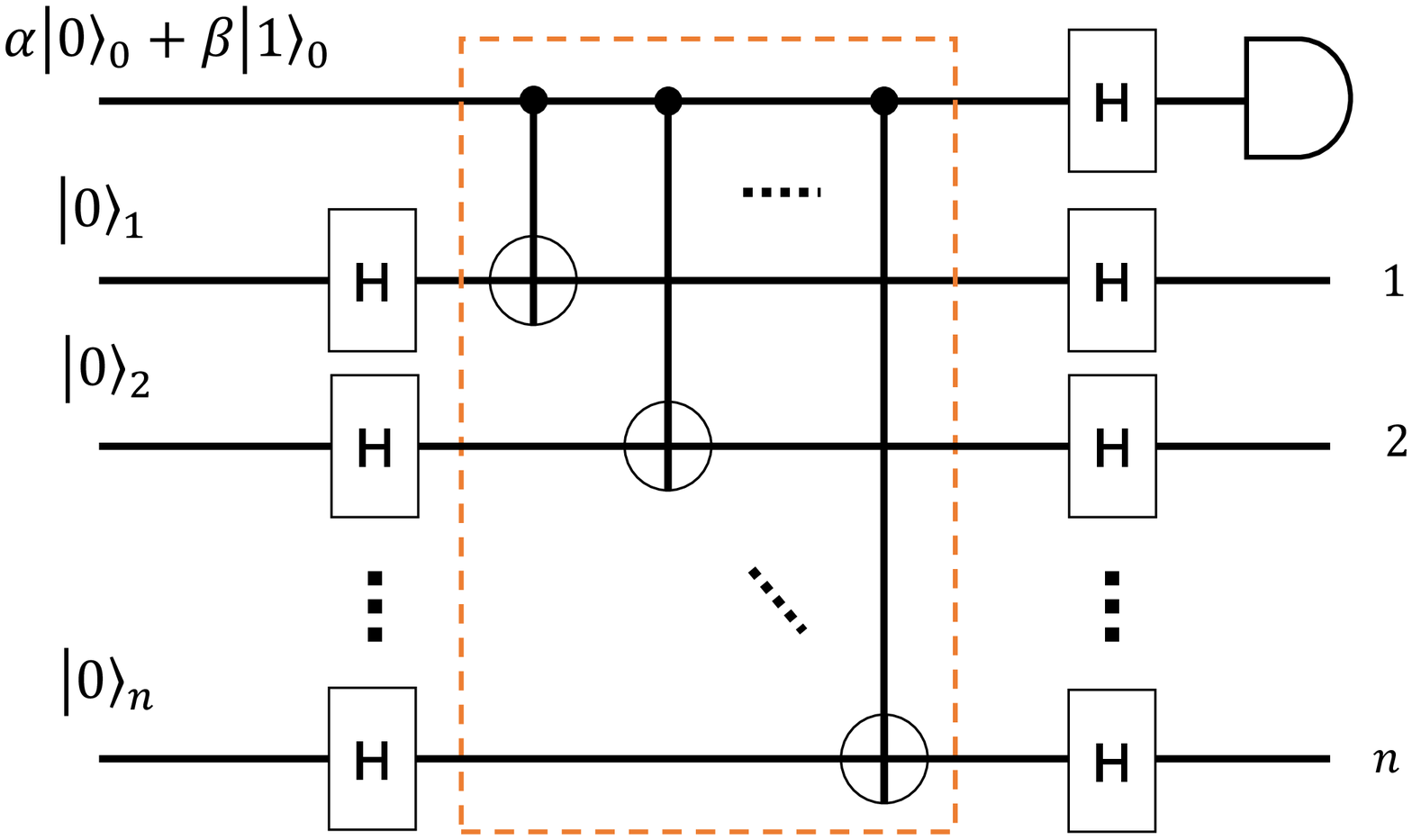}\\
  \caption{Schematic diagram of multi-controlled phase gate and quantum circuit of photon-phonon conveter.}\label{fig6}
\end{figure}
Now we expand our system to a more general model.
As shown in \fref{fig6}, there is an array of quadratically coupled optomechanics, cavity-$k$ ($k>1$) coupled to the cavity-1 with strength $J_{k-1}$.
Each membranae of the optomechanical cavity coupled to a low frequency oscillator with the strength $V_j$. The Hamiltonian of the system can be write as
\begin{eqnarray}
  H &=& \sum_{j=1}^{n}\omega_{j}a_j^{\dag}a_j+\omega_{mj} b_j^{\dag}b_j+\omega_{Aj} b_{Aj}^{\dag}b_{Aj}
  \nonumber\\&&+g_ja_j^{\dag}a_j(b_j^{\dag}+b_j)^2+V_j(b_{j}^{\dag}b_{Aj}+b_{j}b_{Aj}^{\dag})
  \nonumber\\&&+\sum_{s=1}^{n-1}J_{s}(a_1^{\dag}a_{s+1}+a_1a_{s+1}^{\dag}),
\end{eqnarray}
where $a_j$ and $\omega_j$ denotes the cavity photon operator and frequency, respectively. $\omega_{mj}$, $b_j$ and $\omega_{Aj}$, $b_{Aj}$ describe the membranae mode and auxiliary oscillator mode, respectively. The fourth term denotes the optomechanical interaction. The fifth and sixth term denote the interaction between the mechanical modes and
optical modes, respectively.
If we set cavity-1 as an input port of the multi-path convert system, the input photon state can be transmitted from cavity-1 to cavity-$k$ due to the BS(Beam Splitter) interaction.
According to the analysis in former section, we can convert the single-photon state from optical mode to mechanical mode by using cross-Kerr nonlinearity.
Thus, composing the two manipulations, we can convert the arbitrary input single-photon state from cavity-1 to any other auxiliary oscillator through cavity-$k$.
We also notice that the coupling rate $V_j$ can be controlled by adjusting the capacitance $C$ experimentally, and that the key parameter of nonlinear effect $g_{ej}$ is a function of $V_j$.
Thus, it is possible for us to use this system to realize a controllable multi-path single-photon phonon convector.
According to the mode shows in \fref{fig6}, only when the CPF gate is achieved and the optical state is transferred into cavity-$j$ at the time, can we convert the input photon state to the $j$-th phonon state.
In order to evaluate the quality of the conversion, we define the conversion fidelity which is defined as
\begin{eqnarray}
 F_{Cj} &=& \sqrt{F_{Gj}F_{Sj}} \nonumber \\
   &=& (\langle\psi_f| \rho_{j}|\psi_f\rangle\langle\psi_0|\rho_{aj}|\psi_0\rangle)^{1/4},
\end{eqnarray}
where $F_{Gj}$ denotes the fidelity of the CPF gate between $j$-th optical mode and mechanical mode, $\rho_{j}$ is the density operator of them. $\psi_f$ is the final state after a perfect CPF gate operator.
$F_{Sj}$ denotes the fidelity between input state $\psi_0$ and the state in cavity-$j$ $\psi_0$.
To investigate the system, using the same processing in section II, we get the effective Hamiltonian under the condition $\omega_{mj}\gg\{g_{j},V_{j},J_{j}\}$.
\begin{eqnarray}
  H_{eff} &=& \sum_{j=1}^{n}\Delta'_{j}a_j^{\dag}a_j
  +\omega_{ej}b_{Aj}^{\dag}b_{Aj}+g_{ej}a_j^{\dag}a_jb_{Aj}^{\dag}b_{Aj}
  \nonumber\\&&+\sum_{s=1}^{n-1}J_{s}(a_1^{\dag}a_{s+1}+a_1a_{s+1}^{\dag}),
\end{eqnarray}
where $\omega_{ej} = \omega_{Aj}-V_j^2/\omega_{mj}$, $g_{ej}=2g_jV_j^2/\omega_{mj}^2$,
$\Delta_{j}=\omega_{j}-\omega_{jL}$, here $\omega_{jL}$ denotes the driving frequency of cavity-$j$.

\begin{figure}
  \centering
  \includegraphics[width=8cm]{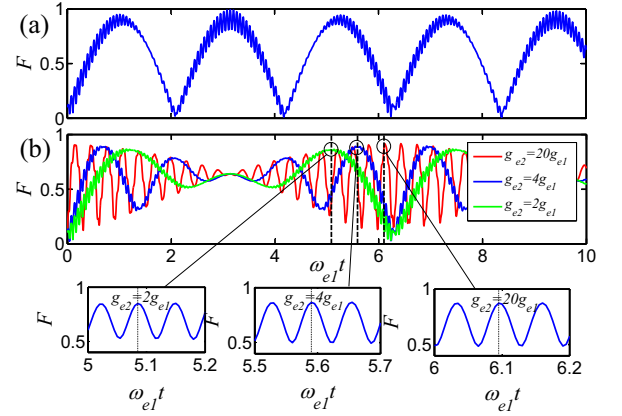}\\
  \caption{(a) The conversion fidelity of output port 1.
  (b) The conversion fidelity of output port 2.
  The bottom of the figure describes the fidelity between input signal and cavity-2.
  Other parameters are $g_{e1}/\omega_{e1}=1$, $\omega_{e1}=\omega_{e2}$, $J_1/\omega_{e1}=0.1$.}\label{fig7}
\end{figure}

\begin{figure}
  \centering
  \includegraphics[width=8cm]{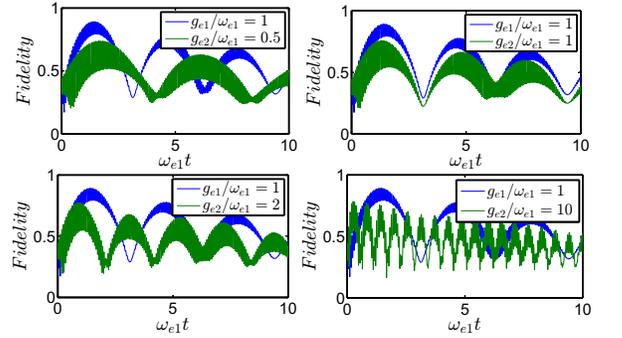}\\
  \caption{The dynamic of the conversion fidelity with different effective coupling rate $g_{e2}$. The blue line denotes the conversion fidelity of output port 1, the green line denotes the conversion fidelity of output port 2.
  Other parameters are $\omega_{e1}=\omega_{e2}$, $\kappa_{1}=\kappa_{2}=0.1\omega_{e1}$, $J_1/\omega_{e1}=0.1$, $\gamma_1=10\gamma_2=10^{-5}\omega_{e1}$, $n_{th1}=n_{th2}=5$.}\label{fig8}
\end{figure}

To set a simple example showing the function of this system, we set $n=2$.
According to the numerical simulation of the fidelity using $H_{eff}$, we show the controllable transmission process of the system.
As shown in \fref{fig7}a, we plot the conversion fidelity of the output port with oscillator-1 without dispassion.
It shows that the CPF gate can be periodically realized at a specific time.
When $t=\pi/g_{eff}$, the CPF gate can be realized with the fidelity $F_{C1}=1$. After implement the single qubit operator, we can transfer the unknown quantum state from optical mode to mechanical mode.
As shown in \fref{fig7}b, we plot the conversion fidelity of the output port with oscillator-2 without dispassion, although the CPF gate can be periodically realized at a specific time during the dynamic of the system, the conversion fidelity $F_{C2}$ can not equal to $1$ due to the effective dumpling from cavity-1.
The realization time of CPF gate is controlled by the value of $g_{eff}/\omega_{eff}$ which can be adjusted by changing the voltage $V_0$ and capacitance $C_0$ experimentally, where $V\propto C_0 V_0$.
To better understand the process of the conversion between input signal and output state with oscillator-2, we plot the fidelity between the input signal and the state in cavity-2 in the bottom of \fref{fig7}.
When the state is transferred from input signal to cavity-2, the signal conversion is realized.
That is to say, just as we have analyzed before, the state converter will be realized only when the input signal is transferred to cavity-2 and the CPF gate is realized between cavity-2 and oscillator-2 in the mean time.
Thus, it is possible for us to establish a controllable state conversion between input signal and output signal from oscillator-(1,2).
Considering the realization, we calculate the conversion fidelity with the dissipation of system.
As shown in \fref{fig8}, we plot the dynamic of conversion fidelity in different effective coupling rate $g_{e2}$.
The converter can be realized periodically, and the output time can be controlled by the effective coupling rate $g_{e2}$.
But the conversion fidelity will decrease due to the accumulation of dissipation in system.
Therefore, it is necessary for us to improve the cavity quality factor to ensure the practicality of the system.

\section{discussion and conclusion}
As is shown in \fref{fig7}, the rest systems can be seen as an environment, which introduces an effective dissipation rate to the sub-system (such as cavity-$j$) we focus on.
Thus, the number of output ports in the multi-path converter is limited to some extent due to the requirement of a high output state fidelity.
To increase the fidelity of converter, on the one hand, we can enlarge the ratio of $|g_{eff}/\omega_{eff}|$ to accelerate the evolution, which will reduce the cumulative effects of system dissipation (shown in \fref{fig3}).
On the other hand, we can directly improve the quality factor of cavity to reduce the effective dissipation rate.
In our scheme, the optomechanical oscillator can be realized by a suspended film bulk acoustic resonator with frequency in the order of $10^{9}Hz$ \cite{Nature.464.697}.
The auxiliary oscillator can be realized by nanoobjects with frequency in the order of $10^{6}Hz$ \cite{nphys.9.480,nature.452.72}.
The interaction between them can be controlled by piezoelectric ceramics and LC circuits \cite{NJP.10.115001,Nature.464.697}.
A two-dimensional electron gas and Schottky-contacted gold electrodes can be used to output the phonon signal we need \cite{PhysRevLett.110.127202}.

We propose a scheme to realize single photon-phonon converter by combining the single-bit operation with the cross-Kerr nonlinear effect between photons and phonons in quadratically coupled optomechanical system.
Considering the realization, we use an auxiliary oscillator to accelerate the evolution of system to improve the fidelity of CPF gate, which achieves $F_{max}\approx 0.97$ under the consideration of dissipation, as is shown in \fref{fig3}.
By controlling the adjustable parameters, we are able to achieve an ultra-strong cross-Kerr nonlinearity ($g_{eff}/\omega_{eff} \gg 1$), which makes sense for the realization of single quantum photon-phonon converter to a large extent.
Moreover, we also extend the converter protocol to a controllable multiple outputs scheme through a dimension extension.
By choosing the detection time of the output signal of cavity-$1$, we can transduce an unknown input single optical state into a mechanical state of an arbitrary output port we want with high fidelity.
This protocol provides us with a possibility to perform photon-phonon multipath conversion and operation.

\appendix
\begin{widetext}
\section*{Appendix:DERIVATION OF THE EFFECTIVE HAMILTONIAN}
Under the weak diving condition and $\omega_{m1}\gg g$, ${b_1}^2$ and ${b_1}^{{\dag}^2}$ can be ignored by rotating wave approximation, the Hamiltonian \eref{hami} can be rewrited as
\begin{eqnarray}
  H_{sys} &=& \omega_c a^{\dag}a-ga^{\dag}a(b_1b_1^{\dag}+b_1^{\dag}b_1)+\sum_{j=1,2}(\omega_{mj}b_j^{\dag}b_j+Vb_{j}^{\dag}b_{3-j})+\varepsilon (a^{\dag}e^{-i\omega_dt}+ae^{i\omega_dt}),
\end{eqnarray}
In a frame rotating at the frequency of optical drive, the nonlinear quantum Langevin equations are given by
\begin{subequations}
\begin{eqnarray}
  \dot{a} &=& -(i\Delta'+\kappa/2)a+2igab_1^{\dag}b_1+\sqrt{\kappa}a_{in}, \\
  \dot{b_{1}} &=& -(i\omega_{m1}+\gamma_{1}/2)b_{1}+2igb_{1}a_1^{\dag}a_1-iVb_2+\sqrt{\gamma_{1}}b_{in,1}, \\
  \dot{b_2} &=& -(i\omega_{m2}+\gamma_2/2)b_2-iVb_1+\sqrt{\gamma_2}b_{in,2},
\end{eqnarray}
\end{subequations}
where $\Delta'_{1}=\omega_c-\omega_{d}-g$ denotes the mechanically modulating detuning of the cavity with a driving frequency $\omega_d$. $\kappa$, $\gamma_1$ and $\gamma_2$ represent the decay rates of mode $a$, $b_1$ and $b_2$.
The cavity's input $a_{in}$ is the sum of coherent amplitudes $\overline{a}_{in}$ and vacuum noise operator $\xi$.
$b_{in,1}$ and $b_{in,2}$ are the noise operators associated with the mechanical dissipations. Defining photon number operator $N_a=a^{\dag}a$ and phtonon number operator $N_b=b_1^{\dag}b_1$, we can rewrite the nonlinear quantum Langevin equations as
\begin{subequations}
\begin{eqnarray}
  \dot{a} &=& -(i\Delta'+\kappa/2)a+2igaN_b+\sqrt{\kappa}a_{in}, \\
  \dot{b_{1}} &=& -(i\omega_{m1}+\gamma_{1}/2)b_{1}+2igb_{1}N_a-iVb_2+\sqrt{\gamma_{1}}b_{in,1}, \\
  \dot{b_2} &=& -(i\omega_{m2}+\gamma_2/2)b_2-iVb_1+\sqrt{\gamma_2}b_{in,2},\\
  \dot{Na}&=&-\kappa_1 N_a,
\end{eqnarray}
\end{subequations}
These equations can be formally integrated as
\begin{subequations}
\begin{eqnarray}
  a_{1}(t) &=& a(0)e^{-(i\Delta'+\kappa/{2})t}e^{\int_0^td\tau 2igN_b(\tau)}+\int_{0}^{t}d\tau e^{-(i\Delta'+\kappa/{2})(t-\tau)}e^{\int_{\tau}^td\tau'2igN_b(\tau')}\sqrt{\kappa}a_{in}(\tau)],\\
  b_{1}(t) &=& b_1(0)e^{-(i\omega_{m1}+\gamma_1/2)t}e^{\int_0^t d\tau 2igN_a(\tau)}\nonumber\\
  &&+\int_0^td\tau e^{-(i\omega_{m1}+\gamma_1/2)(t-\tau)}e^{\int_{\tau}^t d\tau' 2igN_a(\tau')}[-iVb_2(\tau)+\sqrt{\gamma_1}b_{in,1}(\tau)],\\
  b_2(t) &=& b_2(0)e^{-(i\omega_{m2}+\gamma_2/2)t}
  +\int_0^td\tau e^{-(i\omega_{m2}+\gamma_2/2)(t-\tau)}[-iVb_1(\tau)+\sqrt{\gamma_2}b_{in,2}(\tau)].
\end{eqnarray}
\end{subequations}
Under the weak driving condition, the occupation number of the system is rather small. If $\omega_{m1}\gg\{ g,V\}$, we can let $b_1(t) \approx b_1(0)e^{-(i\omega_{m1}+\gamma_1/2)t}$ when calculator the dynamics of $a$ and $b_2$.
Substitute $b_1(t)$ to $a(t)$ and $b_2(t)$. We have
\begin{eqnarray}
  a(t) &\approx& a(0)e^{-(i\Delta'+\kappa_{2}/{2})t}+A_{in}(t), \\
  b_2(t) &\approx& b_2(0)e^{-(i\omega_{m2}+\gamma_2/{2})t}+B_{in,2}(t),
\end{eqnarray}
where $A_{in}(t)\approx\int_{0}^{t}d\tau e^{-(i\Delta'+\kappa/{2})(t-\tau)}\sqrt{\kappa}a_{in}(\tau)$ and $B_{in,2}(t)=\int_0^td\tau e^{-(i\omega_{m2}+\gamma_2/2)(t-\tau)}\sqrt{\gamma_2}b_{in,2}(\tau)$ denote the noise terms.
Substitute $a(t)$ and $b_2(t)$ to $b_{1}(t)$. We have
\begin{eqnarray}
  b_1(t) &\approx& b_1(0)e^{-(i\omega_{m1}+\gamma_1/2)t}e^{\int_0^t d\tau 2igN_a(\tau)}\nonumber\\
  &&+\int_0^td\tau e^{-(i\omega_{m1}+\gamma_1/2)(t-\tau)}e^{\int_{\tau}^t d\tau' 2igN_a(\tau')}[-iVb(0)e^{-(i\omega_{m2}+\gamma_2/{2})\tau}-iVB_{in,2}(\tau)+\sqrt{\gamma_1}b_{in,1}(\tau)],
  \end{eqnarray}
we notice that the Hamiltonian is commutative to operator $N_a$. When the dispassion rate of cavity is weak enough and the dynamic time is much smaller than decoherence time, we can regard $N_a$ as independent of time $t$ in the photon decay period. Under the condition $\omega_{m1}\gg\omega_{m2}$,$\gamma_{1}\gg\gamma_2$ and since the term containing $e^{-\gamma_{1}t}$ is a fast decaying term which can be neglected, we have
\begin{eqnarray}
  b_1(t) &\approx& \frac{-iVb_2(t)}{i\omega_{m1}+\gamma_1/2-2igN_a}+B'_{in,1}
  =-\frac{iVb_2(t)}{(i\omega_{m1}+\gamma_1/2)(1-\frac{2igN_a}{i\omega_{m1}+\gamma_1/2})}+B'_{in,1},
\end{eqnarray}
where the noise term is denoted by $B'_{in,1}(t)\approx\int_0^td\tau e^{-(i\omega_{m1}+\gamma_1/2)(t-\tau)}[-iVB_{in,2}(\tau)+\sqrt{\gamma_1}b_{in,1}(\tau)]$.
Using the equation $\frac{1}{1+(y+ix)} \approx 1-y+(-i+2iy)x$,
and neglecting high level minim when $g\ll\omega_{m1}$, we have
\begin{eqnarray}
  b_1(t) &\approx& \frac{-iVb_{2}(t)}{i\omega_{m1}+\gamma_1/2} \{1+\frac{(-i\omega_{m1}+\gamma_1/2)2gN_b}{A}+O[(\frac{g}{A})^2]\}+B'_{in,1}(t),
\end{eqnarray}
where $A={\omega_{m1}}^2+{\gamma_{1}}^{2}/4$.
Under the condition $\omega_{m1}\gg\gamma_1$, putting $b_1(t)$ back to $\dot{a}$ and $\dot{b_2}$, we finally obtain
\begin{eqnarray}
  \dot{a} &=& -(i\Delta'+\kappa/2)a+ig_{eff}ab_2^{\dag}b_2+\sqrt{\kappa}a_{in},\\
  \dot{b_2}&=&-(i\omega_{eff}+\gamma_{eff}/2)b_2+g_{eff}b_2a^{\dag}a+\sqrt{\gamma_{ef}}B'_{in,2},
\end{eqnarray}
where $\omega_{eff} = \omega_{m2}-\frac{V^2}{\omega_{m1}}$, $g_{eff}=\frac{V^2}{\omega_{m1}^2}2g$, $\gamma_{eff}=\gamma_2+\frac{V^2}{\omega_{m1}^2}\gamma_1$.
Thus, the effective Hamiltonian is
\begin{equation}
H_{eff}=\Delta' a^{\dag}a+\omega_{eff}b_2^{\dag}b_2+g_{eff}a^{\dag}ab_2^{\dag}b_2.
\end{equation}

\end{widetext}

\bibliography{omcv}
\end{document}